\input harvmac
\let\includefigures=\iftrue
\let\useblackboard=\iftrue
\newfam\black

\includefigures
\message{If you do not have epsf.tex (to include figures),}
\message{change the option at the top of the tex file.}
\input epsf
\def\figin{\epsfcheck\figin}\def\figins{\epsfcheck\figins}
\def\epsfcheck{\ifx\epsfbox\UnDeFiNeD
\message{(NO epsf.tex, FIGURES WILL BE IGNORED)}
\gdef\figin##1{\vskip2in}\gdef\figins##1{\hskip.5in}
instead
\else\message{(FIGURES WILL BE INCLUDED)}%
\gdef\figin##1{##1}\gdef\figins##1{##1}\fi}
\def\DefWarn#1{}
\def\figinsert{\goodbreak\midinsert}
\def\ifig#1#2#3{\DefWarn#1\xdef#1{fig.~\the\figno}
\writedef{#1\leftbracket fig.\noexpand~\the\figno}%
\figinsert\figin{\centerline{#3}}\medskip\centerline{\vbox{
\baselineskip12pt\advance\hsize by -1truein
\noindent\footnotefont{\bf Fig.~\the\figno:} #2}}
\endinsert\global\advance\figno by1}
\else
\def\ifig#1#2#3{\xdef#1{fig.~\the\figno}
\writedef{#1\leftbracket fig.\noexpand~\the\figno}%
\global\advance\figno by1} \fi

\def\id{{1 \kern-.28em {\rm l}}}

\def\K3{{\bf K3}}
\def\journal#1&#2(#3){\unskip, \sl #1\ \bf #2 \rm(19#3) }
\def\andjournal#1&#2(#3){\sl #1~\bf #2 \rm (19#3) }

\def\bar{\overline}

\def\ie{{\it i.e.}}
\def\eg{{\it e.g.}}

\def\frac#1#2{{#1\over#2}}

\def\half{\frac12}

\def\inbar{\,\vrule height1.5ex width.4pt depth0pt}
\def\IC{\relax\hbox{$\inbar\kern-.3em{\rm C}$}}
\def\IR{\relax{\rm I\kern-.18em R}}
\def\IP{\relax{\rm I\kern-.18em P}}

%
%

%
\catcode`\@=11
\def\slash#1{\mathord{\mathpalette\c@ncel{#1}}}
\overfullrule=0pt

\def\SS{{\cal S}}

\def\underrel#1\over#2{\mathrel{\mathop{\kern\z@#1}\limits_{#2}}}

\catcode`\@=12


%

\def\det{{\rm det}}

\def\det{{\rm det}}


\lref\WittenZW{
  E.~Witten,
  ``Anti-de Sitter space, thermal phase
   transition, and confinement in  gauge
   theories,''
  Adv.\ Theor.\ Math.\ Phys.\  {\bf 2}, 505 (1998)
  [arXiv:hep-th/9803131].
}

\lref\PeskinEV{
 M.~E.~Peskin and D.~V.~Schroeder,
 {\it An Introduction to quantum field theory,}
 Reading, USA: Addison-Wesley (1995).

entry}

\lref\NambuTP{Y.~Nambu and G.~Jona-Lasinio,
``Dynamical Model Of Elementary Particles Based On An Analogy With
Superconductivity. I,''Phys.\ Rev.\  {\bf 122}, 345 (1961).
}

\lref\SakaiCN{
T.~Sakai and S.~Sugimoto,
``Low energy hadron physics in holographic QCD,''
Prog.\ Theor.\ Phys.\  {\bf 113}, 843 (2005)
[arXiv:hep-th/0412141].
}

\lref\Distler{
  J.~Distler and F.~Zamora,
  ``Chiral symmetry breaking in the AdS/CFT correspondence,''
  JHEP {\bf 0005}, 005 (2000)
  [arXiv:hep-th/9911040].
}

\lref\Klebanov{
  I.~R.~Klebanov and M.~J.~Strassler,
   ``Supergravity and a confining gauge theory: Duality cascades and
  chiSB-resolution of naked singularities,''
  JHEP {\bf 0008}, 052 (2000)
  [arXiv:hep-th/0007191].
}

\lref\BabingtonVM{
  J.~Babington, J.~Erdmenger, N.~J.~Evans, Z.~Guralnik and I.~Kirsch,
  ``Chiral symmetry breaking and pions in non-supersymmetric gauge /
gravity
  duals,''
  Phys.\ Rev.\ D {\bf 69}, 066007 (2004)
  [arXiv:hep-th/0306018].
}

\lref\Evans{
  N.~J.~Evans and J.~P.~Shock,
  ``Chiral dynamics from AdS space,''
  Phys.\ Rev.\ D {\bf 70}, 046002 (2004)
  [arXiv:hep-th/0403279].
}

\lref\Burrington{
  B.~A.~Burrington, J.~T.~Liu, M.~Mahato and L.~A.~Pando Zayas,
   ``Towards supergravity duals of chiral symmetry breaking in Sasaki-Einstein
  cascading quiver theories,''
  JHEP {\bf 0507}, 019 (2005)
  [arXiv:hep-th/0504155].
}

\lref\Bak{
  D.~Bak and H.~U.~Yee,
   ``Separation of spontaneous chiral symmetry breaking and confinement via
  AdS/CFT correspondence,''
  Phys.\ Rev.\ D {\bf 71}, 046003 (2005)
  [arXiv:hep-th/0412170].
}
\lref\GreenDD{
  M.~B.~Green, J.~A.~Harvey and G.~W.~Moore,
  Class.\ Quant.\ Grav.\  {\bf 14}, 47 (1997)
  [arXiv:hep-th/9605033].
}

\lref\KruczenskiBE{
  M.~Kruczenski, D.~Mateos, R.~C.~Myers and D.~J.~Winters,
  JHEP {\bf 0307}, 049 (2003)
  [arXiv:hep-th/0304032].
}

\lref\KruczenskiUQ{
  M.~Kruczenski, D.~Mateos, R.~C.~Myers and D.~J.~Winters,
  JHEP {\bf 0405}, 041 (2004)
  [arXiv:hep-th/0311270].
}

\lref\ItzhakiDD{
N.~Itzhaki, J.~M.~Maldacena, J.~Sonnenschein and S.~Yankielowicz,
``Supergravity and the large N limit of theories with sixteen
supercharges,''
Phys.\ Rev.\ D {\bf 58}, 046004 (1998)
[arXiv:hep-th/9802042].
}

\lref\AntonyanVW{
  E.~Antonyan, J.~A.~Harvey, S.~Jensen and D.~Kutasov,
  ``NJL and QCD from string theory,''
  arXiv:hep-th/0604017.
  }

\lref\GrossJV{
D.~J.~Gross and A.~Neveu,
``Dynamical Symmetry Breaking In Asymptotically Free Field Theories,''
Phys.\ Rev.\ D {\bf 10}, 3235 (1974).
}

\lref\SakaiYT{
T.~Sakai and S.~Sugimoto,
``More on a holographic dual of QCD,''
Prog.\ Theor.\ Phys.\  {\bf 114}, 1083 (2006)
[arXiv:hep-th/0507073].
}

\lref\Witten{
  E.~Witten,
  ``Chiral Symmetry, The 1/N Expansion, And The SU(N) Thirring
Model,''
  Nucl.\ Phys.\ B {\bf 145}, 110 (1978).
}

\lref\Affleck{
  I.~Affleck,
  ``On The Realization Of Chiral Symmetry In (1+1)-Dimensions,''
  Nucl.\ Phys.\ B {\bf 265}, 448 (1986).
}

\lref\MaldacenaCG{
  J.~M.~Maldacena and A.~Strominger,
  ``Semiclassical decay of near-extremal fivebranes,''
  JHEP {\bf 9712}, 008 (1997)
  [arXiv:hep-th/9710014].
}

\lref\AharonyUB{
  O.~Aharony, M.~Berkooz, D.~Kutasov and N.~Seiberg,
  ``Linear dilatons, NS5-branes and holography,''
  JHEP {\bf 9810}, 004 (1998)
  [arXiv:hep-th/9808149].
}

\lref\ItzhakiTU{
 N.~Itzhaki, D.~Kutasov and N.~Seiberg,
 ``I-brane dynamics,''
 JHEP {\bf 0601}, 119 (2006)
 [arXiv:hep-th/0508025].
}

\lref\pw{ A.~M.~Polyakov and P.~B.~Wiegmann, ``Theory Of
Nonabelian Goldstone Bosons In Two Dimensions,'' Phys.\ Lett.\ B
{\bf 131}, 121 (1983).
}

\lref\ghm{ M.~B.~Green, J.~A.~Harvey and G.~W.~Moore, ``I-brane
inflow and anomalous couplings on D-branes,'' Class.\ Quant.\
Grav.\  {\bf 14}, 47 (1997) [arXiv:hep-th/9605033].
}

\lref\DashenXZ{
R.~F.~Dashen, S.~k.~Ma and R.~Rajaraman,
``Finite Temperature Behavior Of A Relativistic Field Theory With
Dynamical
Symmetry Breaking,''
Phys.\ Rev.\ D {\bf 11}, 1499 (1975).
}

\lref\AharonyDA{
O.~Aharony, J.~Sonnenschein and S.~Yankielowicz,
 ``A holographic model of deconfinement and chiral symmetry
restoration,''
arXiv:hep-th/0604161.
}

\lref\ParnachevDN{ A.~Parnachev and D.~A.~Sahakyan, ``Chiral phase
transition from string theory,'' arXiv:hep-th/0604173.
}

\lref\LukyanovNJ{
S.~L.~Lukyanov, E.~S.~Vitchev and A.~B.~Zamolodchikov,
``Integrable model of boundary interaction: The paperclip,''
Nucl.\ Phys.\ B {\bf 683}, 423 (2004)
[arXiv:hep-th/0312168].
}

\lref\LukyanovBF{
S.~L.~Lukyanov and A.~B.~Zamolodchikov,
``Dual form of the paperclip model,''
Nucl.\ Phys.\ B {\bf 744}, 295 (2006)
[arXiv:hep-th/0510145].
}

\lref\GaoUP{
Y.~H.~Gao, W.~S.~Xu and D.~F.~Zeng,
``NGN, QCD(2) and chiral phase transition from string theory,''
arXiv:hep-th/0605138.
}

\lref\AntonyanQY{
  E.~Antonyan, J.~A.~Harvey and D.~Kutasov,
  ``The Gross-Neveu Model from String Theory,''
  arXiv:hep-th/0608149.
}

\lref\KutasovDJ{
D.~Kutasov,
``D-brane dynamics near NS5-branes,''
arXiv:hep-th/0405058.
}

\lref\KutasovRR{
D.~Kutasov,
``Accelerating branes and the string / black hole transition,''
arXiv:hep-th/0509170.
}

\lref\polbooka{
  J.~Polchinski,
  ``String theory. Vol. 1: An introduction to the bosonic string,''
  Cambridge, UK: Univ. Pr. (1998).
}

\lref\polbookb{
  J.~Polchinski,
  ``String theory. Vol. 2: Superstring theory and beyond,''
Cambridge, UK: Univ. Pr. (1998). }

\lref\ahkII{E.~Antonyan, J.~A.~Harvey and D. Kutasov, ``The
Gross-Neveu
model from string theory.''}

\lref\PeetWN{
  A.~W.~Peet and J.~Polchinski,
  ``UV/IR relations in AdS dynamics,''
  Phys.\ Rev.\ D {\bf 59}, 065011 (1999)
  [arXiv:hep-th/9809022].
}

\lref\SkenderisWP{
  K.~Skenderis,
  ``Lecture notes on holographic renormalization,''
  Class.\ Quant.\ Grav.\  {\bf 19}, 5849 (2002)
  [arXiv:hep-th/0209067].
}

\lref\KarchSH{
  A.~Karch and E.~Katz,
  ``Adding flavor to AdS/CFT,''
  JHEP {\bf 0206}, 043 (2002)
  [arXiv:hep-th/0205236].
}

\lref\SonET{
  D.~T.~Son and M.~A.~Stephanov,
  Phys.\ Rev.\ D {\bf 69}, 065020 (2004)
  [arXiv:hep-ph/0304182].
}

\lref\AharonyXN{
  O.~Aharony, A.~Giveon and D.~Kutasov,
  ``LSZ in LST,''
  Nucl.\ Phys.\ B {\bf 691}, 3 (2004)
  [arXiv:hep-th/0404016].
}

\lref\KruczenskiUQ{
  M.~Kruczenski, D.~Mateos, R.~C.~Myers and D.~J.~Winters,
  ``Towards a holographic dual of large-N(c) QCD,''
  JHEP {\bf 0405}, 041 (2004)
  [arXiv:hep-th/0311270].
}

\Title{\vbox{\baselineskip12pt\hbox{EFI-06-18}
\hbox{}}} {\vbox{\centerline{Chiral Symmetry Breaking from}
\bigskip
\centerline{Intersecting D-Branes}
}}
\bigskip

\centerline{\it E. Antonyan, J.~A. Harvey and D. Kutasov}
\bigskip
\centerline{EFI and Department of Physics, University of
Chicago}\centerline{5640 S. Ellis Av. Chicago, IL 60637}

\smallskip

\vglue .3cm

\bigskip

\bigskip
\noindent We study a class of intersecting D-brane models in which
fermions localized at different intersections interact via exchange
of bulk fields. In some cases these interactions lead to dynamical
symmetry breaking and generate a mass for the fermions. We analyze
the conditions under which this happens as one varies the dimensions
of the branes and of the intersections.
\bigskip

\Date{August 2006}

\newsec{Introduction}

The system of $N_c$ $D4$-branes wrapped around a circle with
anti-periodic boundary conditions for the fermions provides an
interesting example of gauge-gravity duality \WittenZW. In a certain
region of the parameter space of the brane configuration,
corresponding to small four-dimensional 't Hooft coupling, the low
energy theory on the D-branes is $(3+1)$-dimensional $SU(N_c)$
Yang-Mills (YM) theory without matter. Unfortunately, in that limit
the theory on the branes is hard to solve, even in the large $N_c$
limit.

For large 't Hooft coupling (and large $N_c$), the dynamics reduces
to supergravity in the near-horizon geometry of the $D4$-branes and
can be analyzed in some detail. The theory exhibits confinement and
has a spectrum of glueballs that can be calculated using
supergravity. In this regime the theory on the branes is not pure
YM, since the adjoint scalars and fermions living on the $D4$-branes
are not decoupled, and their dynamics is not well described by the
effective Lagrangian obtained by dimensional reduction of $N=1$ SYM
in $9+1$ dimensions. The dynamics can be described in terms of the
(2,0) superconformal field theory in $5+1$ dimensions compactified
on a two-torus with twisted boundary conditions around one of the
cycles, but this description is not easy to use.

While the theory with a good supergravity description is not YM, the
two are related by a continuous deformation, and one may hope that
they are in the same phase. The reason for this is that the $(2,0)$
theory is believed to be a standard QFT; compactifying it on a torus
leads to an RG flow, which is expected to be smooth. Changing the
parameters associated with the compactification probes different parts
of this flow. Since physical properties change smoothly along RG trajectories,
it is natural to expect that the supergravity and YM regimes are in the same
phase. This means that some qualitative and perhaps even quantitative 
features of YM theory can be addressed in supergravity. This is of course 
a general theme in gauge-gravity duality.

A natural way to add dynamical fermions in the fundamental
representation of the gauge group to the setup of \WittenZW\ was
proposed in \SakaiCN\ (see \eg\
\refs{\KarchSH\SonET\KruczenskiUQ-\BabingtonVM}, for other work on
incorporating dynamical quarks into gauge-gravity duality). It
involves adding to the $N_c$ color $D4$-branes $N_f$ flavor $D8$ and
$\bar{D8}$-branes, which intersect the color branes along an
$\IR^{3,1}$. The flavor branes and anti-branes are separated by a
distance $L$ in the remaining (compact) direction along the
$D4$-branes. Left-handed quarks live at the intersection of the $D4$
and $D8$-branes, while right-handed quarks live at the $D4-\bar{D8}$
intersection. A nice feature of this brane construction is that
quarks of different chiralities are physically separated in the
extra dimensions, and the chiral $U(N_f)_L\times U(N_f)_R$ flavor
symmetry is manifest.

QCD with massless quarks is obtained in a certain region of the
parameter space of the brane configuration where the dynamics is
again difficult to analyze. In a different region of parameter space
one can analyze the theory by studying the DBI action for the
$D8$-branes in the near-horizon geometry of the $D4$-branes. As in
the case without quarks, in this regime the theory is not QCD but it
is expected to be in the same phase. In particular, it exhibits
confinement and chiral symmetry breaking. The latter has a nice
geometric realization.

It was pointed out in \AntonyanVW\ that the brane construction of
\SakaiCN\ has the interesting property that it decouples the scales
of confinement and chiral symmetry breaking.\foot{A similar
phenomenon was observed in a different context in \Bak.} By varying
the parameters of the brane configuration one can make the energy
scale of chiral symmetry breaking arbitrarily higher than that of
confinement. In fact, sending the size of the circle that the
$D4$-branes wrap to infinity one arrives at a theory which breaks
chiral symmetry but does not confine. This theory can again be
studied at weak coupling using field theoretic techniques and at
strong coupling using supergravity.

The intersecting brane construction of \refs{\SakaiCN,\AntonyanVW}
is a special case of a much more general class of constructions,
obtained by varying the dimensions of the color and flavor branes
and that of the intersection. Some other examples were considered
recently in \refs{\GaoUP,\AntonyanQY}. The purpose of this paper is
to present a more uniform treatment of these and other intersecting
brane systems, using field theory and supergravity, as appropriate,
to analyze them.

The general setup we will consider is the following. We start with
$N_c$ color $Dq$-branes, which generalize the color $D4$-branes of
\refs{\SakaiCN,\AntonyanVW}. We will take these branes to be
non-compact in all directions since we are mainly interested in
chiral symmetry breaking. The low-energy theory on the $Dq$-branes
is $(q+1)$-dimensional SYM with sixteen supercharges. For $q<3$ it is
strongly coupled in the infrared and weakly coupled in the UV, while
for $q>3$ it has the opposite behavior. For $q=3$ this theory is
$N=4$ SYM in $3+1$ dimensions, which is conformally invariant. The
't Hooft coupling of the theory on the color branes,
$\lambda_{q+1}$, has units of $({\rm length})^{q-3}$.

For $q\le 4$ the low energy theory of the color branes is a local QFT,
which can be decoupled from gravity by taking a certain scaling limit.
For $q=5$ it is a non-local theory known as Little String Theory (LST). 
This theory has a Hagedorn density of states; the  Hagedorn temperature 
sets the scale of non-locality (see \eg\ \refs{\AharonyUB,\AharonyXN} 
for discussions). For $q=6$, the theory on the branes cannot be 
decoupled from bulk gravity \ItzhakiDD.

In addition to the color branes we have $N_f$ flavor $Dp$ and
$\bar{Dp}$-branes, which are analogs of the flavor $D8$-branes and
anti-branes in the construction of \refs{\SakaiCN,\AntonyanVW}. The
color and flavor branes intersect along an $(r+1)$-dimensional
spacetime, which we will denote by $Dq \cap Dp = I_r$.  The flavor
branes and anti-branes are separated by a distance $L$ in a
direction along the color branes but transverse to the intersection
$I_r$. In the cases of interest, the light degrees of freedom
localized at the intersections are fermions, which interact via
exchange of fields living on the color branes. 

One can think of the flavor branes, and in particular the dynamics of the 
fermions, as probing the theory of the color branes at the scale $L$. By 
changing $L$ we probe the color theory at different points along its RG 
trajectory. For $q\le 4$ we expect the dependence on $L$ to be smooth. 
For $q=5$ we expect it to be smooth for distances much larger than the 
scale of non-locality of the underlying LST. For $q=6$ there is no 
decoupling limit and one has to take into account gravitational effects.

We will study the dynamics of these configurations in the limit $N_c
\rightarrow \infty$, $g_s \rightarrow 0$ with $g_s N_c$ and $N_f$
held fixed. This dynamics depends non-trivially on the values of
$p$, $q$ and $r$, and on the dimensionless parameter
\eqn\lameff{\lambda_{q+1}^{(\rm eff)}(L)= \lambda_{q+1}L^{3-q}~,}
which can be thought of as the effective coupling of the color
degrees of freedom at distance scale $L$, or energy $E \sim 1/L$. We
will explore the dependence of the low-energy behavior on these
parameters, focusing on the question of dynamical symmetry breaking
at weak and strong coupling.

We have organized the paper as follows. In section 2 we present a
general analysis of the intersecting brane systems we will consider.
We review the spectrum of fields localized at a particular
intersection, and describe the leading interactions at weak coupling
between fields at different intersections. There are two
qualitatively different cases that need to be considered, depending
on whether or not this interaction has finite range.

The supergravity description of the above intersecting brane systems
is obtained by replacing the color branes by their near-horizon geometry
and studying the DBI action of the flavor branes in this background.
Holography suggests that in some cases (depending on the dimension
of the color branes) supergravity provides a useful description of the 
dynamics at strong coupling. 

In the following sections we illustrate the general considerations
of section 2 with a number of specific models. In section 3 we
discuss configurations with color $D4$-branes. These include the
models of \refs{\SakaiCN,\AntonyanVW, \AntonyanQY} as well as a
model with color and flavor $D4$-branes and a $(1+1)$-dimensional
intersection, which we discuss in some detail. In this model there
are left and right-moving fermions with the same quantum numbers
under $U(N_c) \times U(N_f)$ at each intersection, and one can give
them a mass by separating the color and flavor branes in the two
dimensions transverse to both. For zero mass we show that the model
still exhibits the dynamical breaking of a certain chiral symmetry,
both at weak coupling where it is described by a generalization of
the Gross-Neveu model, and at strong coupling where it is described
by probe brane dynamics in gravity. For non-zero mass this symmetry
is explicitly broken. 

In section 4 we discuss models with color $Dq$-branes for $q>4$.
These include a IIB model with $D5$-branes intersecting in $1+1$
dimensions. The weakly coupled theory is the Gross-Neveu model, 
with a slightly different UV cutoff than that of \AntonyanQY. As the
coupling increases (\ie\ as $L$ decreases), one probes shorter and
shorter distance physics in the LST of the fivebranes. As mentioned
above, this theory is non-local. One expects to encounter non-field 
theoretic behavior when $L$ reaches the scale of non-locality. Indeed, 
we find that the supergravity analysis gives in this case a continuous 
set of solutions which exists for some critical value $L=L^*$. These 
solutions appear to describe the interactions of the fermions localized 
at the intersection with the continuum of closed string modes propagating 
in the fivebrane throat. We study a number of additional models with 
color fivebranes, and find that all of them have similar solutions, with 
slightly different values of $L^*$.

We also discuss models with color $D6$-branes. At weak coupling
they can be studied using field theoretic means. At finite coupling
it is not clear that they makes sense due to the absence of a good 
UV completion (which does not involve gravity). The supergravity 
analysis predicts the existence of an unstable state with broken 
symmetry at weak coupling, and is inapplicable for strong coupling.

In section 5 we analyze several models with color $D2$ and
$D3$-branes, and find qualitatively similar behavior to the color
$D4$-brane case. We conclude in section 6 with a discussion.

\newsec{General results}

The brane configurations that we will consider consist of two
intersections, $Dq \cap Dp = I_r$ and $Dq \cap \bar{Dp} = I_r$,
separated by a distance $L$ in a direction transverse to the $Dp$
and $\bar{Dp}$-branes and along the $Dq$-branes. In this section we
develop some tools for analyzing these systems. We start by
reviewing some standard facts about such intersections, following
\polbookb. We then go on to a discussion of systems with two
intersections in a regime where the coupling between them is weak
and the dynamics can be studied using field theoretic techniques. In
the last subsection we describe these systems in supergravity, and
discuss the implications of holography for them.

\subsec{Classification}

Consider an intersection of $N_c$ $Dq$-branes and $N_f$
$Dp$-branes\foot{All the branes here and below are BPS (although the
full brane configurations we will study break all supersymmetry). We
will not discuss intersecting brane systems that contain non-BPS
branes, since the latter have tachyon instabilities that are not of
interest here.} along an $\IR^{r,1}$ (which we refer to as $Dq \cap
Dp = I_r$). We would like to determine the spectrum of massless
states living at the intersection, and in particular its chirality
with respect to the $U(N_c)\times U(N_f)$ symmetry on the branes.

Imagine that all the spatial directions transverse to the
intersection are compactified on circles, so we can apply T-duality
in these directions. The spectrum of massless states in $r+1$
dimensions is invariant under these operations. Hence, we can map
all intersections to a small class of basic ones, and analyze those.
For example, we can T-dualize the $Dp$-branes to $D9$-branes. The
$Dq$-branes turn in the process to $Dr'$-branes, with $r'\ge r$, and
the intersection becomes $Dr' \cap D9 = I_{r'}$.

There are four possibilities, $r'=1,3,5,7$. For $r'=7$, the spectrum
of strings stretched between the color and flavor branes (or $7-9$
strings) contains a NS sector tachyon localized at the intersection.
We will not discuss this case further here. For $r'=5$, the
intersection preserves eight supercharges, and the spectrum of $5-9$
strings contains a massless hypermultiplet in the representation
$(N_c,\bar N_f)$ of the gauge group. For $r'=3$ the massless
spectrum contains a Weyl fermion in the same representation, coming
from the Ramond sector of $3-9$ strings. All the states in the NS
sector are massive. Finally, for $r'=1$ the system again preserves
eight supercharges, which have a particular chirality in the $1+1$
dimensions along the intersection. The massless spectrum consists of
chiral (Weyl) fermions, with the opposite chirality to that of the
supercharges.

Note that in all the above cases, the massless spectrum at the
intersection coming from $r'-9$ strings is chiral. In cases where
$r'=r$, this means that the original intersection, $Dq \cap Dp =
I_r$, also has a chiral spectrum. When $r'>r$, the spectrum at the
original intersection $I_r$ is obtained by dimensionless reduction
from $r'+1$ to $r+1$ dimensions. As is well known, dimensionally
reducing chiral fermions gives non-chiral ones, so the resulting
$(r+1)$-dimensional spectrum is non-chiral.

A closely related fact is that a transverse intersection, \ie\ one
with $q+p-r=9$ or equivalently $r'=r$, has the property that there
are no directions of space transverse to both kinds of branes, so
they always intersect. This is the geometric counterpart of the fact
that one cannot give a gauge invariant mass to chiral fermions.

On the other hand, intersections with $r'>r$ are not transverse, so
the color and flavor branes can be separated in directions
transverse to both. Doing so gives a mass to the fermions at the
intersection which correspond to strings stretching from one brane
to the other. The fact that it is possible to give a gauge invariant
mass to the fermions implies that the latter are not chiral.

It should be noted that the discussion above addressed the question
of chirality with respect to the $U(N_c) \times U(N_f)$ symmetry on
the branes. Even for non-transverse intersections, the fermions may 
be chiral with respect to geometric symmetries from the normal bundle 
to the intersection. The dynamical breaking of such geometric chiral 
symmetries has been studied in the context of AdS/CFT, see {\it e.g.} 
\refs{\Distler\Klebanov\BabingtonVM\Evans-\Burrington}.

\subsec{Weak coupling}

In the previous subsection we discussed the spectrum of states
localized at a given intersection of the form $Dq \cap Dp = I_r$.
Our main interest in this paper is in the systems with two such
intersections described above. The important new feature of such
systems is the interaction between modes localized at the two
intersections.

The leading interaction between the two intersections is due to
exchange of a single color gluon (and, for non-transverse
intersections, scalars as well). This gives rise to a quartic
interaction proportional to the gauge coupling of the color
$Dq$-branes,
\eqn\gaugecoup{g_{q+1}^2 = (2 \pi)^{q-2} g_s {\ell_s}^{q-3}}
where $\ell_s=\sqrt{\alpha'}$ is the string length and $g_s$ is the string coupling.
The $(q+1)$-dimensional
 't Hooft coupling $\lambda_{q+1}$ can be defined in
terms of $g_{q+1}$ as
\eqn\thooftqpo{\lambda_{q+1}={g_{q+1}^2N_c\over(2\pi)^{q-2}}~.}
The quartic interaction due to single gluon exchange is proportional
to
\eqn\fermschem{{\lambda_{q+1}\over N_c}\int d^{r+1}x d^{r+1}y
G_{q+1}(x-y,L) [ \psi_L^{\dag}(x) \cdot \psi_R(y)] [ \psi_R^{\dag}(y)
\cdot\psi_L(x)] }
where $\psi_L$ and $\psi_R$ are fermion fields localized at the two
intersections, respectively, and
\eqn\gdefn{G_{q+1}(x,L)=(x^2+L^2)^{-\half(q-1)}}
is proportional to the $(q+1)$-dimensional massless propagator over
a distance $L$ in the directions along which the flavor branes are
separated and distance $x$ along the intersection.\foot{We have
written the interaction in Euclidean spacetime, which is convenient
for studying the vacuum structure.} Each term in squared brackets in
\fermschem\ is a singlet of global $U(N_c)$, and we suppress the
flavor labels. The expression \fermschem\ is schematic. For any
given intersection one can write it more precisely, as was done for
some cases in \refs{\AntonyanVW,\AntonyanQY} and will be done for
some others below.

The non-local interaction \fermschem\ is non-singular in the UV. One
can think of $L$ as a UV cutoff. The long distance behavior of the
theory depends in an important way on whether this interaction has
finite range or not. For $q-r>2$, the integral
\eqn\localcrit{\int d^{r+1} x G_{q+1}(x,L)}
converges, and one can think of $G_{q+1}$ as an $(r+1)$-dimensional
$\delta$-function smeared over a distance of order $L$. Thus, at
distances much larger than $L$ one can replace \fermschem\ by the
local interaction
\eqn\fermloc{{1\over N_c}\times{\lambda_{q+1}\over L^{q-3}}\times
L^{r-1}\int d^{r+1}x [\psi_L^{\dag}(x) \cdot \psi_R(x)] [
\psi_R^{\dag}(x) \cdot\psi_L(x)]~.}
Each of the factors in front of the integral in \fermloc\ has a
simple interpretation. The first is necessary to get a smooth large
$N_c$ limit; the second is the effective coupling $\lambda^{(\rm
eff)}_{q+1}(L)$ \lameff. The third is a power of the UV cutoff, that
is needed to account for the scaling dimension of the local quartic
coupling. For example, for $r=3$ (\ie\ $(3+1)$-dimensional
intersection) the four-Fermi coupling has dimension six, which means
that one needs a factor of $L^2$ to reach the dimension required of
a Lagrangian. For $r=1$ the third factor in \fermloc\ is absent, in
agreement with the fact that the four-Fermi coupling is in this case
marginal (more precisely marginally relevant).

The theory with a local  interaction \fermloc\ is solvable in the large
$N_c$ limit. For $(1+1)$-dimensional
 intersections, such models exhibit dynamical symmetry
breaking for arbitrarily weak coupling (an example is the
Gross-Neveu model \GrossJV, which appears in the example studied in
\AntonyanQY, and some other brane configurations that will be
mentioned below). For higher dimension $(r>1)$ they typically do not
break chiral symmetry at weak coupling. An example is the original
NJL model \NambuTP, which as we will see appears in string theory as
a low-energy model corresponding to a certain brane configuration.

For $q-r\le 2$ the integral \localcrit\ diverges and the range of
the quartic interaction \fermschem\ is infinite. This makes the
analysis above more subtle and we will leave it to future work.

\subsec{Supergravity analysis}

The discussion of the previous subsection is valid when the effective
coupling \lameff\ is small. For large $\lambda^{(\rm eff)}_{q+1}(L)$ the
interactions between color and flavor degrees of freedom are strong
and one needs to use other tools to analyze them.

The problem without the flavor branes was studied in \ItzhakiDD,
where a qualitatively different behavior was found for $Dq$-branes
with $q \le 4$, and for those with $q=5,6$. In the former case the
theory on the branes can be decoupled from gravity. As one changes
the efective coupling \lameff, the useful description changes from
field theory, to ten-dimensional gravity, and sometimes to
eleven-dimensional gravity. The important fact for our purposes is
that there is a wide range of values of the effective coupling in
which the field theoretic description is strongly coupled and one
has to use type II supergravity to study the dynamics.

For $D5$-branes the situation is more complex. The low-energy field
theory degrees of freedom do not decouple from a continuum of states
that live in the throat of the fivebranes (see \eg\
\refs{\MaldacenaCG,\AharonyUB,\AharonyXN}). Gravity in the
near-horizon geometry of the fivebranes includes these states. For
$D6$-branes, the low-energy theory on the branes cannot be decoupled
from gravity at all.

In this subsection we will analyze what happens when one adds to the
system the flavor branes and anti-branes discussed above. We will
replace the color branes by their near-horizon geometry and will
study the flavor branes and anti-branes using their
Dirac-Born-Infeld (DBI) action. We will see that the results are
compatible with the above picture, and in particular exhibit
qualitatively different behavior for color branes with $q\le4$,
$q=5$ and $q=6$.

We will take the color $Dq$-branes to span the directions $(0,1,
\cdots ,q)$, while the flavor $Dp$ and $\bar{Dp}$-branes are
stretched in $(0,1, \cdots ,r,q+1,q+2, \cdots ,q+p-r)$. The two
intersections lie along the $(r+1)$-dimensional space with coordinates
$(0,1, \cdots r)$ and are separated by a distance $L$ in the $x^q$
direction.

The near-horizon geometry of the $Dq$-branes is described by the
metric and dilaton
\eqn\metcon{\eqalign{ ds^2 = & \left( {U \over R_{q+1}}
\right)^{(7-q)/2}
dx_{||}^2 -
\left( {R_{q+1} \over U} \right)^{(7-q)/2} \left( dU^2 + U^2 d
\Omega_{8-q}^2 \right)~, \cr
e^\Phi = & g_s \left( {R_{q+1} \over U} \right)^{(7-q)(3-q)/4}  ~, \cr
}}
where
\eqn\rpdef{R_{q+1}^{7-q} = (2 \sqrt{\pi})^{5-q} \Gamma \left( {7-q
\over 2} \right) g_s N_c = 2^{7-2q} (\sqrt{\pi})^{9-3q} \Gamma
\left( {7-q \over 2} \right) g_{q+1}^2 N_c~. }
There is also a RR flux through the $(8-q)$-sphere \metcon.

In the supergravity approximation, the dynamics of the flavor
$Dp$-branes in the background \metcon\ is described by a DBI action,
whose form (suppressing the gauge field on the branes) is given by
\eqn\dbidp{S_{Dp}= - T_p \int d^{p+1}x e^{-\Phi} \sqrt{-\det
g_{Dp}}}
where $g_{Dp}$ is the induced metric on the $Dp$-brane. There are
also Chern-Simons couplings in the full action which are important
for the analysis of anomalies, but play no role in what follows.

As in the examples studied in \refs{\AntonyanVW,\AntonyanQY}, in
solving for the shape of the flavor branes we have to allow for the
possibility that the parallel separated $Dp$ and $\bar{Dp}$-brane
configuration that we specified at weak coupling is deformed due to
the effects of interactions with the color branes.

The brane configuration should still approach a brane and anti-brane
at a distance $|\delta x^q|=L$ as $U\to\infty$, and preserve the
same symmetry as the intersecting brane system at weak coupling.
This implies that the $Dp$-brane wraps $\IR^{r,1}$, a spherical
subspace $S^{p-r-1}$ of the $(8-q)$-sphere transverse to the color
branes, and a curve $U(x^q)$ in the $(U,x^q)$ plane, which approaches
$U\to\infty$ as $x^q=\pm L/2$. The induced metric is then given in
terms of $U'=dU/dx^q$ by
\eqn\inddp{\eqalign{ds_p^2 = & \left(U \over R_{q+1}
\right)^{(7-q)/2} \bigl[ \eta_{\mu \nu}dx^\mu dx^\nu \bigr] -
\left(U \over R_{q+1} \right)^{(7-q)/2} \left[1+\left(R_{q+1} \over
U\right)^{7-q} (U')^2 \right] (dx^q)^2 \cr & -  \left( R_{q+1} \over
U \right)^{(7-q)/2} U^2 d \Omega_{p-r-1}^2~.}}
Using \inddp\ and \metcon\ in \dbidp\ leads to the action
\eqn\sdpfin{S_{Dp} = - C(p,q,r) \int dx^q U^{\alpha\over2}
\sqrt{1+\left(U\over R_{q+1}\right)^{2 \beta} (U')^2}}
where we have defined
\eqn\condefGen{\eqalign{ C(p,q,r) & = {T_p \over g_s} {\rm
Vol}(\IR^{r,1}) {\rm Vol}(S^{p-r-1})
R_{q+1}^{{1\over4}(q-7)(2r-p-q+6)} ~, \cr \alpha &=
(2r-q-p+6)\frac{7-q}{2} + 2(p-r-1) ~, \cr
 \beta & = {q-7 \over 2} ~. \cr
                                      }}
For the special case of a transverse intersection these expressions
can be simplified by using the relation $p+q-r=9$.

Since the Lagrangian \sdpfin\ does not depend explicitly on $x^q$,
there is a first integral given by
\eqn\firstint{{U^{\alpha\over2} \over \sqrt{1+\left({U\over
R_{q+1}}\right)^{2 \beta} (U')^2}} = U_0^{\alpha\over2}}
where $U_0$ is the value of $U$ where $U'=0$. Solving \firstint\ for
$U'$ and integrating gives
\eqn\xqeq{x^q(U) = \pm{1\over R_{q+1}^\beta} \int_{U_0}^U {u^\beta
du \over \sqrt{(u/U_0)^\alpha -1}} ~. }
The integral can be evaluated in terms of complete and incomplete Beta
functions,
\eqn\xbeta{x^q(U) = \pm{U_0 \over \alpha} \left({U_0 \over R_{q+1}}
\right)^\beta \left[ B \left(- {\beta+1 \over \alpha} + {1 \over 2},
{1 \over 2} \right) - B \left( \left({U_0 \over U}\right)^\alpha ;
-{\beta+1\over \alpha} + {1 \over 2}, {1 \over 2} \right) \right]
~.}
The boundary conditions $x^q(U\to\infty)\to\pm L/2$ imply that
\eqn\leqn{L= 2 |x^q(\infty)| = {2  U_0 \over \alpha} \left( {U_0
\over R_{q+1}} \right)^\beta B \left(-{\beta+1 \over \alpha}+ {1
\over 2},{1 \over 2} \right) ~. }
Using \thooftqpo, \rpdef, \condefGen\  and  dropping constants this
can be rewritten as
\eqn\oneleq{L^2 \sim  {U_0^{q-5}\lambda_{q+1}}~. }
This is the holographic energy -- distance relation of \PeetWN, with
the field theory energy scale $E \sim 1/L$. The solution \xbeta\
describes a curved, connected $Dp$-brane, which looks like $Dp$ and
$\bar{Dp}$-branes connected by a wormhole whose width is determined
by $U_0$ \leqn.

The case of color fivebranes $(q=5)$ is special: the $U_0$ dependence
in \leqn\ cancels and one finds
\eqn\lfive{L={2\pi R_6\over p-1}~.}
Thus, in this case a solution exists only for a particular $L$ of order $R_6$
and any width $U_0$. The scale \lfive\ is of order the non-locality scale of
the LST on the fivebranes. Thus, it is natural to suspect that it is associated
with interactions between the fermions at the intersections and high energy,
non-field theoretic, excitations in the fivebrane theory.

For a given value of $L$, the equations of motion of the
DBI action \sdpfin\ have two solutions. One corresponds to $U_0=0$ in
\firstint\ and describes a disconnected $Dp$ and $\bar{Dp}$-brane
pair, running along the $U$ axis at $x^q=\pm L/2$. The other is the
curved, connected solution \xbeta. To determine the ground state of
the system, one needs to compare their energies.

The energy difference between the two solutions, $\Delta E \equiv
E_{\rm straight} - E_{\rm curved}$ is
\eqn\endiff{\Delta E = {C(p,q,r) \over R_{q+1}^\beta}  \left(
\int_0^\infty  duu^{\alpha/2+\beta}  - \int_{U_0}^\infty du
{u^{\alpha/2+\beta} \over \sqrt{1-(U_0/u)^\alpha}}  \right) ~. }
Each integral  in \endiff\ is separately divergent at large $u$. The
divergence can be regulated as in \AntonyanVW\ by regrouping terms,
or equivalently by writing the integrals in terms of Beta functions
which are then defined by analytic continuation. This procedure
should be equivalent to the holographic renormalization reviewed in
\SkenderisWP.  One finds
\eqn\eng{\Delta E  = -{1 \over \alpha} {C(p,q,r)
U_0^{\alpha/2+\beta+1} \over R_{q+1}^\beta}B \left(-{1 \over 2} -
{{\beta+1} \over \alpha}, {1 \over 2} \right) ~. }
One can check that the sign of $\Delta E$ depends only on $q$. For
$q\le 4$, $\Delta E>0$, so that the curved solution has lower energy
and is the ground state of the system. For $q=5$, $\Delta E =0$, so
the energy is independent of $U_0$. For $q=6$, $\Delta E <0$; hence,
the curved solution is unstable\foot{For the $D6 \cap D2 = I_1$
intersection, plugging into \condefGen\ one finds $\alpha=0$, so the
present discussion does not apply to this case. The statements in
the text apply to all the other intersections with color
$D6$-branes.} and the straight one is the ground state.

In order to trust the above supergravity analysis the curvature of
the metric must be small at the values of $U$ which govern the
dynamics, that is at $U\sim U_0$. It was shown in \ItzhakiDD\ that
this is the case provided that the effective coupling \lameff\ at
the energy $U_0$ is large,
\eqn\coupcond{\lambda_{q+1} U_0^{q-3} \gg 1~.}
For $q\le 4$, eliminating $U_0$ using \oneleq\ we find that the
validity of supergravity requires
\eqn\regsugra{\lambda_{q+1}\gg L^{q-3}~,}
or, equivalently, large effective coupling at the scale $1/L$. We
see that for supergravity to be valid, the effective coupling should
be large both at the scale $U_0$, and at the scale $1/L$. These 
two scales are dynamically important; the former sets the dynamically
generated mass of the fermions at the intersection, while the latter
governs the mass of the low-lying mesons, which can be studied
by expanding the DBI action around the background solution \xbeta.

The relation between the two scales \oneleq\ can be rewritten as
\eqn\revleq{U_0\sim {1\over L}\left[\lambda_{q+1}^{(\rm
eff)}(L)\right]^{1\over 5-q}\sim {1\over L}\sqrt{\lambda_{q+1}^{(\rm
eff)}(U_0)}~.}
For large effective coupling \coupcond, \regsugra, $U_0$ is a higher
energy scale than $1/L$. Depending on the value of $q$, one of the
conditions \coupcond, \regsugra\ can be more restrictive. For $q=2$
the effective coupling \lameff\ decreases as the energy increases.
Thus, the condition that the coupling be large at energy $U_0$
\coupcond\ is more restrictive than that at $1/L$ \regsugra. For
$q=3$, the coupling does not run and \coupcond\ and \regsugra\ are
equivalent. For $q=4$ the coupling increases with energy so
\regsugra\ is the more restrictive condition.

The cases $q=5$ and $q=6$ have to be treated separately. For 
$q=5$, we see from \lfive\ that the curved solution exists only for a
particular $L$ of order $R_6$. Moreover, since the energy difference 
\eng\ vanishes for this case, we have a continuum of solutions with the 
same energy, labeled by the width of the throat connecting the flavor 
$Dp$-branes, $U_0$. As mentioned above, these solutions are 
associated with the non-locality of the theory of the fivebranes. 

For $q=6$, combining \oneleq\ and \coupcond\ leads to
\eqn\sixcond{\lambda_7\ll L^3~.}
In this case, validity of the supergravity analysis requires that
the effective coupling at the scale $1/L$ be {\it small}, while that
at the scale $U_0$ should still be large \coupcond. Note that the
two requirements are consistent since $U_0\gg 1/L$ and the coupling
on the $D6$-branes increases with energy.

At first sight it is surprising that the supergravity approximation
should be valid at weak coupling, where we already have a good
description of the dynamics in terms of a low-energy field theory.
This is a manifestation of the non-decoupling of the $D6$-branes
from bulk gravity. The curved brane solution we found in this case
describes the interactions of the fermions with gravity, and it is
not surprising that it is unstable. The interactions between the
fermions and the field theoretic degrees of freedom living on the
color branes take place at much smaller values of $U$, and cannot be
described using supergravity (see \PeetWN\ for related comments).

To summarize, we are led to a natural generalization of the picture
in \ItzhakiDD\ to the system with probe D-branes. For $q\le 4$, the
supergravity analysis describes the strong coupling behavior of the
intersecting brane system. For $q=5,6$ it instead describes the
interactions of the fields associated with the intersection with
non-field theoretic degrees of freedom, the LST modes living in 
the throat of the fivebranes for $q=5$, and gravity modes for $q=6$.

\newsec{$D4-D4$ system}

Our main purpose in the rest of this paper is to illustrate the
general considerations of section 2 in some examples. We have
arranged the discussion by the dimension of the color branes. In
this section we focus on color $D4$-branes; in the next two we
discuss systems with higher and lower-dimensional color branes
respectively.

Two models with color $D4$-branes were discussed in detail in
\refs{\SakaiCN,\AntonyanVW,\AntonyanQY}. In both the intersections
were transverse. In this section we will study one additional
example, with a non-transverse intersection. The color and flavor
branes are in this case both $D4$-branes, and are oriented as
follows:
\eqn\ddconfig{\eqalign{\qquad & 0 ~~~ 1 ~~~ 2 ~~~ 3 ~~~ 4 ~~~5 ~~~ 6
~~~ 7 ~~~ 8 ~~~ 9 ~~~ \cr D4_c: ~~& {\rm x} ~~~ {\rm x} ~~~{\rm x}
~~~ {\rm x} ~~~ {\rm x} ~~~{} ~~~ {} ~~~{} ~~~ {} ~~~ {} ~~~ \cr
D4_f, ~~ \bar{D4}_f: ~~& {\rm x} ~~~ {\rm x} ~~~ {} ~~~ {}~~~ {} ~~~
{}~~~~~ {\rm x} ~~~ {\rm x} ~~~ {\rm x} ~~~ {} ~~~ {} ~~~ \cr }}
The color  $D4$-branes are stretched in $(01234)$ and located at the
origin in $(56789)$. The flavor $D4$ and $\bar{D4}$-branes are
stretched in $(01567)$ and separated by a distance $L$ in $(234)$.
We will take the separation to be in the $x^4$ direction and study
the dynamics in the directions common to the different branes,
$(01)$.

The subgroup of the Lorentz group preserved by this brane
configuration is \eqn\breaklor{SO(1,1)_{01}\times SO(2)_{23}\times
SO(3)_{567}\times SO(2)_{89}} Further global symmetry arises from
the gauge symmetry on the flavor $D4$-branes,
\eqn\gaugesym{U(N_f)_{D4}\times U(N_f)_{\bar{D4}}~.} Comparing to
the discussion of subsection 2.1, we see that since the number of DN
directions for strings stretched between the color and flavor branes
is equal to six, and this number is invariant under T-duality, this
system is T-dual to $D3-D9$. Therefore $r'=3$, while the dimension
of the intersection is $r=1$. Indeed, the intersection is not
transverse as the color and flavor branes can be separated in the
directions $(89)$, which are orthogonal to both. This deformation gives
mass to the fermions at each intersection and breaks the $SO(2)_{89}$
symmetry \breaklor.

The low-energy degrees of freedom in this case are open strings
stretched between color branes, which give rise to $(4+1)$-dimensional
SYM theory with sixteen supercharges, and strings stretched between
color and flavor branes, which give spacetime fermions. To see how
these fermions transform under the global symmetries, one can
proceed as follows. If all the branes in \ddconfig\ were extended in
the (89) directions instead of being localized in them, the
$SO(1,1)_{01}\times SO(2)_{89}$ symmetry in \breaklor\ would have
been extended to $SO(1,3)_{0189}$. The spectrum at each intersection
would then be the same as in \refs{\SakaiCN,\AntonyanVW}, \ie\ a
left-handed Weyl fermion, $q_L$, at one intersection, and a
right-handed one, $q_R$, at the other.

The configuration \ddconfig\ can be obtained by compactifying
$(x^8,x^9)$ on a torus, applying T-duality in both directions and
decompactifying back, in the process breaking the $SO(1,3)$ symmetry
back to $SO(1,1)\times SO(2)$. The massless states at each
intersection are invariant under this operation, so all we have to
do is decompose the left and right-handed $SO(1,3)_{0189}$ spinors
$q_L$, $q_R$ under $SO(1,1)_{01}\times SO(2)_{89}$:
\eqn\onea{q_L = \pmatrix{\chi_{L+} \cr \chi_{R-} \cr };\qquad
                 q_R= \pmatrix{\psi_{R+} \cr \psi_{L-}}~.}
$\chi$ and $\psi$ denote fermions localized at the two
intersections. The subscripts $(L,R)$ and $(+,-)$ on the right hand
sides keep track of chirality in $(01)$ and $(89)$, respectively.
For example, $\chi_{L+}$ in \onea\ is a complex, left-moving (one
component) spinor field in $1+1$ dimensions, with charge $+1/2$
(\ie\ half that of a vector) under $SO(2)_{89}$. Its adjoint,
$\chi_{L+}^*$, is a left-moving fermion with the opposite
$SO(2)_{89}$ charge. Note that unlike the $(3+1)$-dimensional system
discussed in \refs{\SakaiCN,\AntonyanVW} and the $(1+1)$-dimensional
one of \AntonyanQY, here there are left and right-handed fermions at
each intersection. Thus, the two $U(N_f)$ factors in \gaugesym\ no
longer act purely on left and right-handed fermions.

In addition to their Lorentz charges, the fermions \onea\ transform
in the fundamental $(N_c)$ representation of the color gauge group
$U(N_c)$. Under the global symmetry \gaugesym, the fermions
$\chi_{L+}$, $\chi_{R-}$ transform as $(\bar N_f,1)$, while
$\psi_{R+}$, $\psi_{L-}$ transform as $(1,\bar N_f)$.

Much of the discussion of the $D4-D6$ system in \AntonyanQY\ carries
through with little change. In particular, for $L\gg\lambda$ the
color degrees of freedom are weakly coupled, and the dynamics of the
fermions is described by a Lagrangian of the form
\eqn\lreff{\eqalign{ \SS_{\rm eff}= & i\int d^2x
\left(q_L^\dagger\bar\sigma^\mu\partial_\mu q_L
+q_R^\dagger\sigma^\mu\partial_\mu q_R\right) \cr & ~~~ +{g_5^2
\over 4 \pi^2} \int d^2x d^2y G_5(x-y,L) \left( q^\dagger_L(x)\cdot
q_R(y) \right) \left( q^\dagger_R(y)\cdot q_L(x) \right) \cr }}
which is obtained by integrating out the color gauge field in the
single gluon approximation. $G_5(x,L)$ is the five-dimensional
massless propagator, \gdefn.

The $SU(N_c)$ singlet fermion bilinear that enters the four Fermi
interaction \lreff\ can be expressed in terms of two-dimensional
fermions as follows:
\eqn\onec{q^\dagger_L(x)\cdot q_R(y)=\chi_{L+}^*(x) \cdot\psi_{R+}(y)
+
 \chi_{R-}^*(x)\cdot \psi_{L-}(y) ~.}
The resulting four-Fermi interaction is not equivalent to a Thirring
model for $U(N_c)$. This is a direct consequence of the fact that
the color gluons that are exchanged by the fermions at the two
intersections and give rise to \lreff\ include vectors and scalars
under the $(1+1)$-dimensional Lorentz group associated with the
intersection.

Nevertheless, the theory can be solved at large $N_c$ using standard
methods, as in \AntonyanQY. The fermion bilinear \onec\  develops
dynamically a non-zero vacuum expectation value. This breaks
\eqn\breaksym{U(N_f)_{D4}\times U(N_f)_{\bar{D4}}\to U(N_f)_{\rm
diag}~.} Despite appearances, this symmetry breaking is chiral.
Indeed, defining $Q_1$ and $Q_2$ to be the $U(1)$ generators in
$U(N_f)_{D4}$ and $U(N_f)_{\bar{D4}}$, respectively, and $R$ to be
the generator of $SO(2)_{89}$, the combination\foot{Note that $Q_5$
is a linear combination of a symmetry that is broken by \breaksym,
$Q_1-Q_2$, and one that is preserved, $R$.} \eqn\qqff{Q_5 = Q_1 -
Q_2 + 2R} acts chirally on the fermions. The left-moving fermions
$\chi_{L+}$, $\psi_{L-}$ have charge $+2$ and $-2$ respectively,
while the right-moving fermions are neutral. The symmetry \qqff\ is
preserved by the action \lreff, and (if it is a symmetry of the
vacuum) prevents the generation of a mass for the quarks \onea. The
quark bilinear \onec\ has charge $-2$ under it. Thus, if it develops
an expectation value, the symmetry is broken and a quark mass can be
generated.

A very similar analysis to that of \AntonyanQY\ shows that the
expectation value \onec\ takes again the form
\eqn\poscond{\langle q_L^\dagger(x)\cdot
q_R(0)\rangle=N_cm_f\int_{|k|<\Lambda}
{d^2k\over(2\pi)^2}{e^{ik\cdot x}\over k^2+m_f^2} }
with the dynamically generated mass $m_f$ given by
\eqn\mmff{m_f\simeq\Lambda e^{-{L\over\lambda_4}}~.} $\Lambda$ is
the UV cutoff of the theory, $\Lambda\simeq 1/L$. In this case, one
can also analyze the system in the presence of mass terms in the
Lagrangian which explicitly break the chiral symmetry \qqff,
\eqn\massbreak{\delta\CL_{\rm
eff}=m_1\chi_{L+}^*\chi_{R-}+m_2\psi_{R+}^*\psi_{L-} +{\rm c.c.}}
corresponding to separating the color and flavor branes in the (89)
plane. This leads to a straightforward generalization of the
analysis in \AntonyanQY.

As discussed in section 2, at strong coupling the dynamics of the
fermions and color degrees of freedom can be described by studying
the DBI action of the flavor $D4$-branes in the near-horizon
geometry of the color branes, \sdpfin.

The U-shaped solution in which the flavor branes and anti-branes are
connected by a wormhole is given by \xbeta,
\eqn\ushaped{x^4(U) =\pm {1 \over 4} {R_5^{3/2} \over U_0^{1/2}}
\left[ B\left({5\over8},{1\over2}\right) - B\left({U_0^4\over U^4};
{5\over8},{1\over2}\right) \right]~.}
The energy difference between the straight brane and anti-brane
configuration, and the U-shaped one \ushaped\ is given by \eng,
\eqn\endiff{\Delta E= -\frac{1}{4}C(4,4,1)U_0^{3\over2}
B\left(-{3\over8},{1\over2}\right) \approx 0.225\,
C(4,4,1)U_0^{3\over2}~.}
Thus, the vacuum of the theory breaks chiral symmetry both for weak
coupling and for strong coupling, in agreement with the general
analysis of section 2. One can also analyze the system for finite
temperature, as was done for the $D4-D8$ case in
\refs{\AharonyDA,\ParnachevDN} and for the $D4-D6$ case in
\AntonyanQY.

Overall, we conclude that the $D4-D4$ system behaves in a very
similar way to the $D4-D6$ one analyzed in \AntonyanQY. At weak
coupling it reduces to a GN-type model which can be analyzed using
field theoretic techniques, and at strong coupling it can be studied
using the DBI action for the flavor branes in the near-horizon
geometry of the color ones. One advantage of this system is that one
can turn on current masses to the fermions and study the dynamics as
a function of these masses. Another advantage is that the $D4-D4$
brane configuration is simple to lift to M-theory, where the color
$D4$-brane background goes over to $AdS_7\times S^4$, and the flavor
$D4$-branes and anti-branes become  $M5$-branes in this background.
These and other issues deserve further study.

\newsec{Higher-dimensional color branes}

In section 2 we saw that when the color branes are higher than
four-dimensional, the supergravity analysis exhibits some qualitative
differences from the case when their dimension is four or less. In
this section we will examine some examples with color $D5$ and
$D6$-branes to study these phenomena in more detail.

\subsec{Color fivebranes}

In this subsection we discuss a few intersecting brane systems in
which the color branes are $(5+1)$-dimensional. They can be further
subdivided by the type of flavor branes, the dimension of the
intersection and the range of the non-local four-Fermi interaction
at weak coupling. In table 1 we list the four systems that will be
discussed below.

\bigskip
\vbox{
$$\vbox{\offinterlineskip
 \hrule height 1.1pt \halign{&\vrule width 1.1pt#
 &\strut\quad#\hfil\quad& \vrule width 1.1pt#
 &\strut\quad#\hfil\quad& \vrule width 1.1pt#
 &\strut\quad#\hfil\quad& \vrule width 1.1pt#
 &\strut\quad#\hfil\quad& \vrule width 1.1pt#\cr
 &\hfil flavor branes& &\hfil dimension of intersection& &\hfil range
of interaction& \cr
 height0pt &\omit& &\omit& &\omit& \cr
 \noalign{\hrule} height0pt &\omit& &\omit& &\omit& \cr
 &\hfil $D5$& &\hfil $1+1$& &\hfil $L$& \cr
 height0pt &\omit& &\omit& &\omit& \cr
 \noalign{\hrule} height0pt &\omit& &\omit& &\omit& \cr
 &\hfil $D3$& &\hfil $1+1$& &\hfil $L$&  \cr
 height0pt &\omit& &\omit& &\omit& \cr
 \noalign{\hrule} height0pt &\omit& &\omit& &\omit& \cr
 &\hfil $D5$& &\hfil $3+1$& &\hfil $\infty$&  \cr
 height0pt &\omit& &\omit& &\omit& \cr
 \noalign{\hrule} height0pt &\omit& &\omit& &\omit& \cr
 &\hfil $D7$& &\hfil $3+1$& &\hfil $\infty$&  \cr
 height0pt &\omit& &\omit& &\omit& \cr
 }
 \hrule height 1.1pt }$$
} \centerline{\rm Table 1: Different systems with color $D5$-branes
that are discussed in this section.}
\bigskip
\noindent Our first example is obtained by T-dualizing the $D4-D6$
system discussed in \AntonyanQY . It contains color $D5$-branes
stretched in (012345), transversally intersecting flavor $D5$ and
$\bar{D5}$-branes stretched in (016789). A single intersection of
this sort was studied in \refs{\ghm,\ItzhakiTU}. The main new
phenomenon here is the attractive interactions between the fermions
at the two intersections.

As before, we can try to analyze this system using QFT techniques at
weak coupling, and supergravity at strong coupling. The weak
coupling analysis is quite analogous to that of \AntonyanQY . The
fermions at the two intersections, $q_L$, $q_R$, are chiral. Their
dynamics is governed by the effective action \lreff, with the
coupling $g_5$ replaced by $g_6$ \gaugecoup\ and the Green function
$G_5(x,L)$ replaced by $G_6(x,L)$, \gdefn. It is integrable,
\eqn\intsixd{\int d^2x G_6(x,L)={\pi\over 2L^2}~.}
Thus, at distances much larger than $L$ the system reduces to the GN
model with action
\eqn\lgn{ \SS_{gn}= \int d^2x
\left[iq_L^\dagger\bar\sigma^\mu\partial_\mu q_L
+iq_R^\dagger\sigma^\mu\partial_\mu q_R +{\lambda_{gn}\over N_c}
\left( q^\dagger_L(x)\cdot q_R(x) \right) \left( q^\dagger_R(x)\cdot
q_L(x) \right) \right]}
and coupling
\eqn\gnsix{\lambda_{gn}={\pi\lambda_6\over 2L^2}~.}
In particular, it dynamically breaks chiral symmetry  and generates
a fermion condensate \poscond, which leads to the fermion mass
\eqn\mmferm{m_f \simeq {1\over L} e^{-2 \pi/\lambda_{gn}}~.}
As before, this analysis is reliable for $\lambda_6\ll L^2$ and breaks
down when this condition is violated. The DBI analysis does give a
solution in which the flavor $D5$ and $\bar{D5}$-branes are
connected by a wormhole whose shape is given by \xbeta,
\eqn\xfouru{x^5(U) =\pm {1 \over 4} R_6 \left[
B\left({1\over2},{1\over2}\right) - B\left({U_0^4\over U^4};
{1\over2},{1\over2}\right) \right]~.}
However, the asymptotic separation of the $D5$ and $\bar{D5}$-branes
\lfive\ is fixed,
\eqn\lnr{L=\half\pi R_6~.}
For this value of $L$ there are solutions with arbitrary width
$U_0$, whose energy is independent of $U_0$. This is different from
the situation in systems with color $D4$-branes, where there is a
solution for generic $L$, and the width of the wormhole $U_0$ is a
function of $L$, growing when $L$ decreases.

What does this mean for the dynamics of the fermions living at the
two intersections? For $L\gg R_6$ the field theoretic GN analysis is
valid. Chiral symmetry is dynamically broken, and the fermions get a
mass \mmferm. As $L$ decreases, the coupling \gnsix\ grows, and the
dynamically generated mass \mmferm\ does as well. The system 
probes higher and higher energies in the fivebrane theory. As $L$ 
approaches the value \lfive\ the dynamics becomes dominated by high 
energy LST states. 

The resulting physics is not field theoretic and its analysis is beyond the scope 
of this paper. One expects non-smooth behavior at the non-locality
scale \lnr, and it is not clear whether the system exists for
smaller $L$. The problem is analogous to the analysis of fivebrane
thermodynamics, with the inverse temperature $\beta$ being the analog
of $L$ and the energy density on the fivebranes an analog of the fermion
mass $m_f$. The curved solution \xfouru\ is an analog of the Euclidean
continuation of the non-extremal fivebrane solution. The latter has the
property that the circumference of Euclidean time at infinity, $\beta$, is independent
of the energy density, just like in \xfouru\ the asymptotic separation between
the two arms of the U-shape, $L$, is independent of the fermion mass (or $U_0$). 

In the case of fivebrane thermodynamics it is believed that the Euclidean
black hole solution is not continuously connected to the low temperature 
thermodynamics (see \AharonyXN\ for a discussion). It would be interesting
to understand whether in our case the solution \xfouru\ is continuously related
to the large $L$ regime, and what happens for $L$ smaller than \lnr. 

We next move on to the brane system on the second line in table 1,
which is obtained by T-duality from the one discussed in section 3.
It contains color $D5$-branes stretched in (012345) and flavor $D3$
and $\bar{D3}$-branes stretched in (0167).

At weak coupling this system reduces to a GN-type model of the sort
discussed in section 3. In particular, it exhibits dynamical
symmetry breaking. At strong coupling one needs to analyze the DBI
action \sdpfin, which leads to the brane profile
\eqn\shapeuu{x^5(U) =\pm {1 \over 2} R_6 \left[
B\left({1\over2},{1\over2}\right) - B\left({U_0^2\over U^2};
{1\over2},{1\over2}\right) \right]~.}
This looks very similar to the $D5-D5$ solution \xfouru. As there,
the asymptotic separation \lfive\ is fixed, $L=\pi R_6$. Here too we
expect chiral symmetry breaking for sufficiently small values of the 
coupling $\lambda_6/L^2$, and non-smooth behavior when the 
coupling approaches the critical value \lfive.

The last two lines in Table 1 correspond to systems with $(3+1)$-dimensional
intersections. In one, the flavor branes are $D5$ and
$\bar{D5}$-branes stretched in (012367). In the other, they are $D7$
and $\bar{D7}$-branes stretched in (01236789).

In the model with flavor $D5$-branes, each intersection preserves
eight supercharges. Hence the fermions living at a given
intersection belong to a hypermultiplet. SUSY is completely broken
in the full system and exchange of fields living on the color branes
leads to an attractive interaction between the hypermultiplets
localized at the two intersections. At weak coupling, this
attractive interaction has a structure similar to \lreff, however
unlike the system analyzed in \AntonyanQY\ and those discussed
earlier in this paper, the Green function $G_6(x)$ is not integrable
in this case: \eqn\intgg{\int d^4xG_6(x)=\int{d^4x\over
(x^2+L^2)^2}} is logarithmically divergent, so the attractive
interaction has infinite range. Such systems are in general more
subtle than those with a short-range interaction. We hope to return
to their study in a separate publication.

In the supergravity approximation, the solution of the equations of
motion of the DBI action \sdpfin\ is again given by \xfouru, and we
find that there is a curved brane solution in the supergravity
regime for $L$ given by \lnr. The interpretation is the same as there.

The situation is similar for the last brane configuration in table
1, which has flavor $D7$-branes and a $(3+1)$-dimensional intersection
with the color $D5$-branes. This model is T-dual to the $D4\cap D8 =
I_3$ system studied in \AntonyanVW. Thus the spectrum is the same as
there: a left-handed fermion $q_L$ in the $(N_c,\bar N_f,1)$ of
$U(N_c)\times U(N_f)_L\times U(N_f)_R$ at the $D5-D7$ intersection,
and a right-handed fermion $q_R$  in the $(N_c,1,\bar N_f)$ at the
$D5-\bar{D7}$ one. The leading single gluon exchange interaction
between  the left and right-handed fermions takes a form similar to
\lreff\ (or, more precisely, eq. (3.5) in \AntonyanVW). As in the
previous example, this leads to a long-range interaction.

In the supergravity approximation, the solution to the DBI equations
of motion is
\eqn\shapeuu{x^5(U) =\pm {1 \over 6} R_6 \left[
B\left({1\over2},{1\over2}\right) - B\left({U_0^6\over U^6};
{1\over2},{1\over2}\right) \right]~.}
It again has a fixed value of $L$ \lfive\ at which the dynamics
becomes dominated by highly excited LST states.

Before leaving the case of color fivebranes we would like to point
out that the curved brane solutions we found for this case are closely
related\foot{In fact for the $D5 \cap D3 = I_1$ system they are S-dual.}  to the
hairpin brane of \refs{\LukyanovNJ,\LukyanovBF} which plays a role
in analyzing the dynamics of D-branes propagating in the vicinity of
$NS5$-branes \refs{\KutasovDJ,\KutasovRR}. They can be thought of as
continuations to Euclidean space of accelerating brane solutions.
This also makes it plausible that, as mentioned above, they
owe their existence to the interactions of the fields at the
intersections with the continuum of modes living in the fivebrane
throat, \ie\ to Little String Theory dynamics
\refs{\AharonyUB,\AharonyXN}.

\subsec{Color sixbranes}

As mentioned above, due to lack of decoupling, it is not clear whether
intersecting brane systems involving color $D6$-branes make sense
beyond the field theory approximation (\ie\ for finite $L$). Nevertheless,
in this subsection we will discuss two examples of such systems 
using the tools outlined in section 2.

The first system  is obtained by reversing the roles of the $D4$ and 
$D6$-branes in the configuration studied in \AntonyanQY.  Thus, we 
have $N_c$ color $D6$-branes stretched in (0156789) and $N_f$ 
flavor $D4$ and $\bar{D4}$-branes stretched in (01234) and separated 
by a distance $L$ in (56789). At weak coupling the low-energy dynamics 
is again governed by a GN model which breaks chiral symmetry and 
generates a mass for the fermions. The GN coupling $\lambda_{gn}$ 
\lgn\ is proportional to $\lambda_7/L^3$ as can be verified by integrating 
the $(6+1)$-dimensional propagator \gdefn.

{}From our experience with the $D5$-brane case, we expect the
supergravity analysis to be more subtle. The DBI action leads in
this case to the solution
\eqn\xfiveu{x^6(U) = \pm{1\over 3} \sqrt{R_7U_0} \left[
B\left({1\over3},{1\over2}\right) - B\left({U_0^3\over U^3};
{1\over3},{1\over2}\right) \right]~.}
The asymptotic separation between the flavor branes and anti-branes
is
\eqn\lsixfour{L={2\over3}\sqrt{R_7U_0}B\left({1\over3},{1\over2}\right)~.}

Thus, unlike the fivebrane case, here there is a solution in which
the flavor branes are connected by a wormhole for generic $L$.

As discussed in section 2, the supergravity analysis is valid at
large $L$, \sixcond. In that region the dynamics of the fermions is
described by the GN model. The curved brane solution \xfiveu\ is not
a consequence of that dynamics. Instead, its existence is due to
gravitational interactions of the fermions in the vicinity of the
$D6$-branes. This dynamics should be unimportant at low
energies. Indeed, the solution \xfiveu\ is unstable. The energy
difference \eng\ is given by
\eqn\ensixfour{\Delta {E}  \sim  -\frac{1}{3}U_0^{2}
B(-{2\over3},{1\over2}) \approx -0.351\, U_0^{2}~.}
One can show that the curved brane configuration is unstable to
perturbations of the form $U(x^6)\to U(x^6) +\delta U$ with $\delta
U$ independent of $x^6$. Thus, in the supergravity approximation the
vacuum corresponds to straight branes. As mentioned above, the GN
analysis implies that the flavor branes do curve towards each other
and connect, but this happens at a much smaller value of $U$ than
\lsixfour\ and is well outside the regime of validity of
supergravity.

We do not have a good description of chiral symmetry breaking in the 
strongly coupled regime $\lambda\gg L^3$. As mentioned above, it is 
likely that this is because the system does not exist in that regime,
or more generally for any finite $\lambda/L^3$.

Our second example consists of $D6$ and $\bar{D6}$-branes with a
$(3+1)$-dimensional intersection (\ie\ $D6 \cap D6 = I_3$). The color
branes can be taken to lie in (0123456), while the flavor branes and
anti-branes are stretched in (0123789) and separated by a distance
$L$ in $x^6$. This configuration is T-dual to the $D4-D8-\bar{D8}$
one studied in \refs{\SakaiCN,\AntonyanVW}. The spectrum contains
left-handed fermions at one intersection, and right-handed ones at
the other.

At weak coupling these fermions interact via the four-Fermi
interaction given in eq. (3.5) in \AntonyanVW. However, here this
interaction is local, since
\eqn\integr{\int d^4x G_7(x,L)=\int {d^4x\over (x^2+L^2)^{5\over2}}}
is finite. Therefore, the model reduces at long distances $x\gg L$
to a local NJL model \NambuTP , with coupling $\lambda_7/L$ which
has dimension length squared. This model does not break chiral
symmetry at arbitrarily weak coupling. Hence the same should be true
for the intersecting D-brane system in the limit $L^3\gg\lambda_7$.

The supergravity analysis leads to results that are qualitatively
similar to the previous case. The DBI action \sdpfin\ leads to a
solution corresponding to flavor branes and anti-branes connected by
a wormhole
\eqn\solsixsix{ x^6(U) =\pm {1\over 4} \sqrt{R_7U_0}\left[
B({3\over8},{1\over2}) - B({U_0^4\over U^4}; {3\over8},{1\over2})
\right]~.}
This solution is again valid for large $L$, \sixcond, and is unstable,
\eqn\desixsix{ \Delta {E}  \sim -\frac{1}{4}U_0^{5\over 2}
B(-{5\over8},{1\over2}) \approx -0.193\, U_0^{5\over 2}~.}

\newsec{Lower-dimensional color branes}

In this section we will discuss intersecting brane configurations
with $D2$ and $D3$ color branes. The low-energy theories on these
branes are renormalizable gauge theories which can be decoupled from
gravity. For $D3$-branes this theory is $N=4$ SYM. It is conformal,
and its effective coupling \lameff\ is independent of the separation
$L$. For $D2$-branes the theory is weakly coupled in the UV and
strongly coupled in the IR. Thus, the effective coupling \lameff\
grows as $L$ increases.

Due to the low dimension of the color branes, all our examples
involve $(1+1)$-dimensional intersections. Thus, the flavor branes are
codimension one or two defects in the gauge theory. The single gluon
exchange interaction \fermschem\ relevant for the weak coupling
analysis is a long range one, since the integral \localcrit\
diverges. We will not discuss the weakly coupled theory in detail
here.

\subsec{Color threebranes}

Our first example contains $N_c$ color $D3$-branes stretched in
$(0123)$ and $N_f$ flavor $D7$ and $\bar{D7}$-branes stretched in
$(01456789)$ separated by the distance $L$ in $x^3$. This
configuration is T-dual to the $D4-D6$ one considered in
\AntonyanQY. Therefore, it has the same massless spectrum of
fermions at the intersections -- a left-handed fermion $q_L$
transforming in the $(N_c,\bar{N_f},1)$ of $U(N_c)\times
U(N_f)_L\times U(N_f)_R$ at the $D3-D7$ intersection and a
right-handed fermion $q_R$ in the $(N_c,1,\bar{N_f})$ at the
$D3-\bar{D7}$ intersection.

At weak coupling the leading interaction  of the fermions is a long
range GN coupling due to single gluon exchange. At strong coupling,
the $D3$-branes are replaced by their near-horizon geometry,
$AdS_5\times S^5$, and the $D7$-branes are described by the DBI
action \sdpfin. The solution \xbeta\ takes in this case the form
\eqn\xthreeseven{x^3(U)=\pm{1 \over 6} {R_4^{2} \over U_0} \left[
    B({2\over3},{1\over2}) - B({U_0^6\over U^6}; {2\over3},{1\over2})
\right]~. }
The distance between the $D7$ and $\bar{D7}$-branes is
\eqn\distseven{L={R^2\over3U_0}B({2\over3},{1\over2})~. }
One can check using \eng\ that this solution has lower energy than
the one in which the flavor branes are stretched in $U$ at fixed
values of $x^3$. Thus, at strong coupling ($\lambda_4\gg1$) the
chiral symmetry is dynamically broken, and the fermions get a mass
of order $\sqrt{\lambda_4}/L$.

Replacing the flavor branes by $D5$-branes extended in $(014567)$
leads to a T-dual of the $D4-D4$ system studied in section 3.
Therefore, the spectrum is the same as there. It consists of
fermions $\chi_{L+}$, $\chi_{R-}$ transforming in $(N_c,\bar N_f,1)$
and $\psi_{R+}$, $\psi_{L-}$ transforming in $(N_c,1,\bar N_f)$ of
$U(N_c)\times U(N_f)_{D5}\times U(N_f)_{\bar{D5}}$. At weak coupling
there are long range interactions between the fermions, the leading
of which is given by \lreff, with $G_5$ replaced by $G_4$. At strong
coupling, the DBI action of the fivebranes is proportional to that
of the  $D7$-branes discussed above. Thus, the solution is given
again by \xthreeseven, \distseven\ and chiral symmetry is broken at
strong coupling.

\subsec{Color twobranes}

A system with color $D2$-branes and flavor $D8$ and
$\bar{D8}$-branes was considered recently in \GaoUP. The $D2$-branes
can be taken to lie in the directions $(012)$, while the $D8$-branes
span the directions $(013456789)$. This system is T-dual to the
$D3-D7$ one described above, so the massless spectrum at the
intersection is the same as there. At weak coupling (which in this
case means small $L$) it reduces to a non-local GN model, while at
strong coupling (large $L$) it breaks chiral symmetry, as follows
from the DBI analysis of section 2.3.

Replacing the $D8$-branes by $D6$-branes stretched in the directions
$(0134567)$ one finds a $D2-D6$ system T-dual to the $D4-D4$ one of
section 3. At strong coupling, the DBI analysis of section 2 leads
to the solution
\eqn\curveddsix{ x^2(U) = {1 \over 8} {R_3^{5\over2} \over
U_0^{3\over2}} \left[ B({11\over16},{1\over2}) - B({U_0^8\over U^8};
{11\over16},{1\over2}) \right]~.}
This solution breaks chiral symmetry and has lower energy than the
symmetry preserving one, as in all other cases with $Dq$ color
branes with $q\le 4$.

\newsec{Conclusions}

In this paper we presented an analysis of dynamical symmetry
breaking in a class of intersecting $D$-brane systems which
generalize those investigated earlier in 
\refs{\SakaiCN,\AntonyanVW,\GaoUP,\AntonyanQY}. At weak 
coupling these models are usefully classified
by the range of the effective interaction between fermions localized
at the intersections. The models with short range interactions are
straightforward to analyze. When the intersection is
$(1+1)$-dimensional, they exhibit symmetry breaking of the type
studied by Gross and Neveu \GrossJV. In $3+1$ dimensions they reduce
to variants of the Nambu-Jona-Lasinio model \NambuTP\ and do not
lead to symmetry breaking at weak coupling.

We also described the dynamics of these systems in the approximation
where we replace the color branes by their near-horizon geometry and
study the Dirac-Born-Infeld action of the flavor branes in this
geometry. As expected from \refs{\ItzhakiDD,\PeetWN}, the results of
this analysis depend on the dimensionality of the color branes. For
$Dq$ color branes with $q \le 4$ it provides a holographic
description of the corresponding field theory at strong coupling.
For all such systems the ground state exhibits dynamical symmetry
breaking.

For $q>4$, the DBI analysis does not describe the strong coupling
behavior of the field theory at the intersection, but rather the
interaction of modes associated with the intersection with other,
non-field theoretic degrees of freedom. For $q=5$ these are Little
String Theory modes that propagate in the throat of the fivebranes.
Their interactions with the modes at the intersections lead to the
existence of flavor brane configurations in the near-horizon geometry, 
all of the same energy, which are labeled by the width of the throat, 
$U_0$. The asymptotic separation of the branes has a particular value,
\lfive, for these solutions. We interpreted this value as the non-locality
scale of LST for these probes. For $q=6$ the modes in
question are gravity modes of the full theory. Their interaction
with the fermions at the intersection leads to the existence of
unstable configurations of the flavor branes which exhibit dynamical
symmetry breaking.

The main gap in our discussion is the weak-coupling analysis of
intersecting brane systems with long-range fermion interactions.
These systems are subtle, but also potentially important as they
arise in embeddings of QCD in string theory in the limit where the
scale of chiral symmetry breaking is much larger than that of
confinement \AntonyanVW.  We hope to return to them elsewhere.

One of the main motivations for this work was to see whether
intersecting brane configurations of the sort described in
\refs{\SakaiCN,\AntonyanVW,\GaoUP,\AntonyanQY} and in this paper, are
always in the same phase as far as dynamical symmetry breaking is
concerned (in the limit where the color branes are non-compact, so
there is no confinement). In all cases where we were able to analyze 
the dynamics for both weak and strong coupling we found that there was 
no phase transition between the weak and strong coupling regimes
for $Dq$ color branes with $q\le 4$. It is natural to 
conjecture that this is always the case. A better understanding of 
weakly coupled systems with long range interactions would allow us 
to test this conjecture in a wider class of models.

For $q=5$ one does not expect such smoothness to extend beyond the
non-locality scale of the underlying LST, and indeed we found signs
of this in the existence of curved branes solutions for a particular
value of $L$, \lfive. It would be interesting to understand the physics
of these solutions in LST better. For $q=6$ it is not clear  that the brane 
configurations in question exist for finite $L$, so the question of smoothness 
in $L$ does not arise.

\bigskip\medskip\noindent
{\bf Acknowledgements:} We thank O. Aharony, N. Itzhaki and K.
Skenderis for discussions. JH and DK thank the Aspen Center for
Physics for providing a supportive atmosphere during the completion
of this work. The work of EA and JH  was supported in part by NSF
Grant No. PHY-00506630. DK was supported in part by DOE grant
DE-FG02-90ER40560. This work was also supported in part by the
National Science Foundation under Grant 0529954. Any opinions,
findings, and conclusions or recommendations expressed in this
material are those of the authors and do not necessarily reflect the
views of the National Science Foundation.

\listrefs
\end